\documentstyle[preprint,aps,prl]{revtex}
\begin{document}


\preprint{hep-th/9805146, UPR-798-T, ITP/NSF-98-065}
\title{Microstates of Four-Dimensional Rotating Black Holes 
from Near-Horizon Geometry}
\author{Mirjam Cveti\v{c}$^{(1,2)}$ and Finn Larsen$^{(1)}$}
\address{\small 
(1) David Rittenhouse Laboratories, University of Pennsylvania,
Philadelphia, PA 19104 \\
\small (2) Institute for Theoretical Physics, University of California,
Santa Barbara, CA 93106}


\maketitle
\begin{abstract}
We show that a class of four-dimensional rotating black holes 
allow five-dimensional embeddings as black rotating strings. 
Their near-horizon geometry factorizes locally as a product of 
the three-dimensional anti-deSitter space-time and a two-dimensional 
sphere ($AdS_3 \times S^2$), with angular momentum encoded in the 
global space-time structure. Following the observation that the 
isometries on the $AdS_3$ space induce a two-dimensional 
(super)conformal field theory on the boundary, we reproduce the 
microscopic entropy with the correct dependence on the black hole 
angular momentum.  
\end{abstract}
\newcommand {\bea}{\begin{eqnarray}}
\newcommand {\eea}{\end{eqnarray}}
\newcommand {\be}{\begin{equation}}
\newcommand {\ee}{\end{equation}}

Recent developments in nonperturbative string theory have provided 
a fruitful framework to consider quantum properties of black holes. 
In particular, extreme black holes with Ramond-Ramond (R-R) 
charges can be interpreted in higher dimensions as intersecting 
D-branes (the nonperturbative objects in string theory that carry such 
charges~\cite{polch95}), and this has lead to a counting of black hole 
quantum states that agrees precisely with the Bekenstein-Hawking (BH) 
entropy~\cite{strom96a}. This counting is carried out in the weakly 
coupled regime where the D-brane constituents of the black hole 
experience {\it flat} space-time geometry; however, due to 
supersymmetry, it remains valid in the regime where the D-branes 
are strongly coupled, and the geometric space-time description of 
black holes emerges. Thus the microscopic derivation of the 
BH-entropy is justified, but it is difficult to explore the 
quantum black hole geometry in detail using D-branes.

The success of the D-brane counting overshadowed prior
attempts to shed light on the microscopics of black holes in
string theory. In pioneering work, Sen attempted 
to identify the microstates of extreme electrically charged black
holes  with perturbative excitations of string theory~\cite{sen95}. 
However, it was not until the 
discovery of extreme dyonic black holes in string theory --- 
with regular horizons and thus finite BH-entropy --- 
that a quantitative agreement between the microscopic and macroscopic 
entropy became feasible~\footnote{Such black holes were 
originally specified by four Neveu-Schwarz Neveu-Schwarz (NS-NS)
charges~\cite{dyon}, two electric and two magnetic ones. Note, 
however, that these can be mapped onto solutions with  R-R charges, 
by exploiting duality symmetry; their space-time is thus the same.}. 
The microscopic features of these black holes are captured by string 
theory in the {\it curved} space-time geometry specified by their 
near-horizon region~\cite{structure,cfthair1}. In particular, 
a $SL(2,{\bf Z}) \times SU(2)$
Wess-Zumino-Witten (WZW) model~\cite{lowe94,cfthair1} reproduces, 
at least qualitatively~\cite{cfthair1,cfthair2}, the extreme black 
hole entropy directly from the near-horizon geometry.

However, it is only very recently that a precise derivation of the 
BH-entropy from the near-horizon geometry was achieved by 
Strominger~\cite{btzentropy} (and also by 
Sachs {\it et al.}~\cite{dublinbtz}). 
The central observation is that, when embedded in a higher dimensional 
space, the near-horizon geometry locally contains the three-dimensional 
anti-deSitter space-time ($AdS_3$), whose quantum states are specified 
by a two-dimensional conformal field theory (CFT) on the boundary.
This sets the stage for a remarkably robust microscopic counting, which 
precisely reproduces the BH-entropy. The result has
stirred a renewed interest in addressing the details of
black hole microscopics, as they emerge from features of the black 
hole near-horizon geometry~\cite{strom98a,martinecads3}.

The most recent approach reproduces the BH-entropy of static near-extreme 
black holes in five~\cite{btzentropy} and four~\cite{bl98} space-time 
dimensions. However, the black hole solutions of the 
effective low-energy
string theory include as special cases the familiar black holes of 
Maxwell-Einstein gravity; in particular, the four-dimensional 
(neutral rotating) Kerr black hole that is believed to be of 
astrophysical significance, and the (charged rotating) Kerr-Newman 
black hole. It is thus important to generalize the method to more 
general backgrounds, including {\it rotating} black holes.
   
In this paper we address the microscopics of {\it near-extreme rotating
black holes in four dimensions} by exploring their near-horizon region; 
and we elucidate the role of angular momentum in the microscopic 
description. We interpret these black holes as charged rotating 
strings in five dimensions, with the string wrapped around the 
additional dimension.
In M-theory this configuration corresponds to the intersection of a 
rotating configuration of three intersecting M5-branes with momentum 
along their one common direction, identified with the string
in five dimensions. In the decoupling limit the geometry 
is locally a direct product of the three-dimensional anti-deSitter 
space-time and a two-dimensional sphere ($AdS_3 \times S^2$); this 
factorized structure is obtained after a transformation of the angular 
coordinates into a ``co-moving coordinate system''. By identifying the 
microstates with those of the CFT induced on the boundary of the $AdS_3$ 
we reproduce the BH-entropy~\footnote{Related work on five-dimensional 
rotating black holes is presented in  detail elsewhere~\cite{cl98a}.}.

The starting point is a large class of four-dimensional black holes  
(of toroidally compactified string theory), whose explicit space-time 
metric is given in~\cite{cy96b}. They are specified by their mass $M$, 
four $U(1)$ charges $Q_i$ and the angular momentum $J$ or,
more conveniently, in terms of the non-extremality parameter $m$, 
four boosts $\delta_i$ and the angular parameter 
$l$~\footnote{The notation follows~\cite{cy96b}. The $r_0$ 
of~\cite{hlm} is $r_0=2m$, the $\mu$ of~\cite{cl97b} is $m=4\mu$, 
the $l$ of~\cite{cl97b} is $l_{\rm here}=4l_{\rm there}$, the $Q_i$ 
of~\cite{bl98} is $Q_{\rm here}=2Q_{\rm there}$.}: \bea
G_4 M &=& {1\over 4}m\sum_{i=0}^3 \cosh 2\delta_i~, \nonumber \\
G_4 Q_i &=& {1\over 4}m\sinh 2\delta_i~~~;~~i=0,1,2,3~,\nonumber \\
G_4 J &=& ml(\prod_{i=0}^3 \cosh\delta_i - \prod_{i=0}^3 \sinh\delta_i)~,
\nonumber 
\eea
where $G_4$ is the four-dimensional Newton's constant.
The Kerr-Newman black hole corresponds to the case where the 
four charges are identical.
The extreme limit is obtained by
taking, $m\rightarrow 0$ and $l\rightarrow 0$ while keeping 
$Q_{0,1,2,3}$ finite; in this case  $J=0$. From the explicit 
solution one finds the BH-entropy~\cite{cy96b}:
\be
S\equiv {A_4\over 4G_4} =  {\pi\over 4G_4}~
[ 8m^2 (\prod^3_{i=0}\cosh\delta_i+\prod^3_{i=0}\sinh\delta_i)
+
8m\sqrt{m^2-l^2} 
(\prod^3_{i=0}\cosh\delta_i-\prod^3_{i=0}\sinh\delta_i)]~,
\label{eq:macroent}
\ee
where $A_4$ is the area of the outer horizon.

A specific representation of the metric and its accompanying
matter fields is given in~\cite{cy96b} in terms of the NS-NS fields, 
{\it e.g.}, its higher-dimensional interpretation is that of a 
rotating fundamental string with winding and momentum modes, 
superimposed with the Kaluza-Klein
monopole and the H-monopole~\cite{cfthair0}. A particular
duality transformation leaves the four-dimensional space-time 
invariant, while mapping
this configuration to three intersecting $M5$-branes of $M$-theory 
(specified by $Q_{1,2,3}$), with momentum (specified by $Q_0$) 
along the common string. This 
M-theory configuration can be interpreted as a rotating string in 
five dimensions after toroidal compactification. The space-time metric 
of the rotating string is rather complicated, and we were unable to 
write it in a relatively compact form~\footnote{The complications 
associated with the angular momentum are similar to those of adding an 
additional charge (the ``fifth parameter'') to the 
configuration~\cite{cygeneral}.}. However, the metric simplifies 
significantly in the near-horizon region $r\ll Q_{1,2,3}$, when 
the dilute gas  condition 
$\delta_{1,2,3}\gg 1$ is satisfied. Then the metric of the 
five-dimensional rotating string in the Einstein frame becomes:
\bea
ds^2_5 &=& {2\over \lambda}
\left[ (r-{l^2\over 2m}\cos^2\theta) (-d{\tilde t}^2 + d{\tilde y}^2)
+ 2m (1 - {l^2\over 2m^2})\cos^2\theta
d{\tilde t}^2 -{l^2\over m}\cos^2\theta
d{\tilde t}d{\tilde y}\right]\nonumber  \\
&+& {\lambda^2\over 4} \left[ {1\over r^2 -2mr +l^2}dr^2 + d\theta^2
+\sin^2\theta d\phi^2\right]
-\sqrt{\lambda l^2\over m}~( d{\tilde y}+d{\tilde t})
\sin^2 \theta d\phi~,
\label{5drs}
\nonumber \eea
where the boosted variables (specifying the momentum along the string)  
are:
\be
d{\tilde t} = \cosh\delta_0 dt - \sinh\delta_0 dy~,~~~ 
d{\tilde y} = \cosh\delta_0 dy - \sinh\delta_0 dt~,
\nonumber 
\ee
and the characteristic length scale $\lambda$ is defined as 
$
\lambda \equiv (Q_1 Q_2 Q_3)^{1\over 3}$. Note that the metric
(\ref{5drs}) retains nontrivial dependence on the angular momentum;
however, the Kerr-Newman black holes are {\it not} compatible
with the limit considered here.

Introducing the shifted coordinate:
\be
d{\tilde\phi} = d\phi - {2l\over\sqrt{\lambda^3 m}}
 (d{\tilde y}+d{\tilde t})~,
\label{eq:shift}
\ee
yields the factorized metric:
\bea
ds^2_5 &=& {2\over \lambda}\left[-(r-2m+{l^2\over 2m})d{\tilde t}^2
-{l^2\over m}d{\tilde t}d{\tilde y} + 
(r-{l^2\over 2m})d{\tilde y}^2\right]
+\nonumber \\
&+& {\lambda^2\over 4}  \left[ {dr^2\over r^2 -2mr +l^2}+
d\theta^2+\sin^2\theta d{\tilde\phi}^2 \right]~ . 
\nonumber
\eea
With this choice of coordinates it is apparent that the
geometry is a direct product of a two-sphere $S^2$, with radius 
${\lambda\over 2}$, and  a Banados, Teitelboim and Zanelli (BTZ)
black hole in three space-time dimensions with a negative cosmological 
constant $\Lambda=-\lambda^2$. Indeed, introducing 
the coordinates: 
$\tau\equiv {{t\lambda}\over R_{11}}$,~$\sigma\equiv {y\over R_{11}}$, 
and $\rho^2\equiv{2R_{11}^2\over \lambda}
[r + 2m\sinh^2\delta_0 - {l^2\over 2m}(\cosh\delta_0-\sinh\delta_0)^2]~,
$
where $R_{11}$ is the radius of the dimension wrapped by the
string, we find the 
standard BTZ metric~\cite{btz}: 
\bea
ds^2_5 &=& - N^2 d\tau^2 + N^{-2}d\rho^2 + \rho^2 
(d\sigma - N_\sigma d\tau)^2 + 
{1\over 4}\lambda^2 d{\tilde\Omega}^2_3 ~,\nonumber \\
N^2 &=& {\rho^2\over \lambda^2} - M_3
+ {16G_3^2 J_3^2\over \rho^2}~,~~~N_\sigma = {4G_3 J_3\over \rho^2}~,
\nonumber
\eea
where the effective BTZ mass $M_3$ and 
angular momentum $J_3$ are:  
\bea
M_3 &=& 
{R^2_{11}\over\lambda^3} \left[ (4m - {2l^2\over m})\cosh 2\delta_0 +
{2l^2\over m} \sinh 2\delta_0 \right]~,\nonumber \\
8G_3 J_3 &=& {R^2_{11}\over \lambda^2} 
\left[{2l^2\over m} \cosh 2\delta_0
+(4m - {2l^2\over m})\sinh 2\delta_0 \right]~,
\nonumber 
\eea
and the  effective three-dimensional gravitational coupling $G_3$
is related to the four-dimensional one  $G_4$ as~\cite{bl98}:
\be
{1\over G_3} = {1\over G_4}~{A_2\over 2\pi R_{11}}=
{1\over G_4}~{\lambda^2\over 2R_{11}},
\nonumber 
\ee
where $A_2$ is the area of the two-sphere $S^2$. The BTZ geometry 
is {\it locally} $AdS_3$ but global identifications
ensure causal structures that are similar to those familiar from 
four-dimensional black holes. For our purposes it is crucial that
the geometry is {\it asymptotically} $AdS_3$, because then the 
isometries induce a CFT on the boundary 
at infinity~\cite{adsc,btzentropy}. Its central charge $c$ is determined 
by the effective cosmological constant $-\lambda^2$  as~\cite{adsc}:
\be
c = {3\lambda\over 2G_3} = 6~{Q_1 Q_2 Q_3\over 8G_4 R_{11}}~,
\label{eq:ccharge}
\ee
and the conformal weights $h_{L,R}$ (eigenvalues of the Virasoro 
operators $L_0$, ${\bar L}_0$, respectively) are related to the 
BTZ parameters as:
\be
h_{L,R} = {\lambda M_3\pm 8G_3J_3\over 16G_3}~.
\label{eq:cdim}
\ee
Collecting the formulae (\ref{eq:ccharge}) and (\ref{eq:cdim})
we find, in the semi-classical regime where the conformal weights 
are large, the statistical entropy:
\be
S= 2\pi(\sqrt{{c\over 6} h_L}+ \sqrt{{c\over 6} h_R})
= {\pi\over 4G_4}~\sqrt{Q_1 Q_2 Q_3}\left[\sqrt{m}~e^{\delta_0}
+\sqrt{m-{l^2\over m}}~e^{-\delta_0}\right]~.
\label{eq:microent}
\ee
On the other hand, in the dilute gas limit, i.e. $\delta_{1,2,3}\gg 1$, 
the BH-entropy (\ref{eq:macroent}) becomes:
\be
S\simeq  {\pi\over 4G_4}~\sqrt{Q_1 Q_2 Q_3}
\left[\sqrt{m}~e^{\delta_0}+\sqrt{m-{l^2\over m}}~e^{-\delta_0}\right]~.
\nonumber 
\ee
This is in  precise agreement with the
microscopic calculation (\ref{eq:microent}). 
It also agrees with the D-brane motivated counting given in~\cite{hlm}.

The derivation of statistical black hole entropy does not rely on the 
details of the underlying quantum theory, but the relation to 
M-theory is interesting. In M-theory units $R_{11}=g\sqrt{\alpha^\prime}$,
the Planck length is $l_{p}=(2\pi g)^{1\over 3}\sqrt{\alpha^\prime}$,
and 
$G_4 = {1\over 8}~{ (\alpha^\prime)^4 g^2\over R_1 R_2 R_3 R_4 R_5R_6}~$ 
where the $R_i$ are the radii of the compact dimensions and $g$
is the string coupling constant.

In the above we assumed the near-horizon approximation
$r\ll Q_{1,2,3}$ and the dilute gas limit $\delta_{1,2,3}\gg 1$.
These become exact in the formal decoupling limit~\cite{juanads}:
\be
(l_{p}, \ r,\ m, \ l) \to 0\ ,  \ \ {\rm with}\ \  
(r\sim l_{p}^3, \ m\sim l_{p}^3, \  l\sim l_{p}^3, \
R_{1,\cdots , 6}\sim l_p ,\ R_{11}\sim 1)
\ , \label{eq:dec}
\nonumber 
\ee
where the field theory on the intersection of the M-branes decouples from 
gravity. Note, in particular, that angular momentum is compatible with 
decoupling. This appears only to be the case for configurations 
that correspond to regular black holes in four and five 
dimensions; the near-horizon geometry of, {\it e.g.}, the 
$D3$-brane and the $M5$-brane do not have rotating versions. 
Thus only the induced CFTs in {\it two} 
dimensions seems to have world-volume currents with charges that 
can be interpreted as angular momenta.

The quantization conditions on the D-brane charges are~\cite{polch95}
$Q_i ={1\over 2\pi}~ {(2\pi\sqrt{\alpha^\prime})^3\over 
R_{2i-1}R_{2i}} n_i g $, where $n_{1,2,3}$ is the number of 
coincident M5-branes with a given orientation, so
$Q_1 Q_2 Q_3 = n_1 n_2 n_3~8G_4 R_{11}~$; and from (\ref{eq:ccharge}) 
the quantized form of the central charge $c$ becomes $c=6n_1 n_2 n_3$ 
as expected~\cite{strom96d,kt,bl96}. A heuristic microscopic 
interpretation of this formula is that each of the M-branes traverse 
the intersection string $n_{i}$ times, giving a total of $n_1 n_2 n_3$ 
distinct topological sectors, each associated with $6$ degrees of 
freedom.

The quantum numbers $\epsilon$ and $p$ for the string 
energy and momentum, respectively, are introduced
through:
\be
E  = 2m\cosh 2\delta_0 = 8G_4~{\epsilon\over R_{11}}~,~~~
Q_0= 2m\sinh 2\delta_0 = 8G_4~{p\over R_{11}}~,~~~
\nonumber 
\ee
and then the conformal weights $h_{L,R}$ can be written as:
\be
h_{L} = {R_{11}\over 8G_4}~me^{2\delta_0}={1\over 2}(\epsilon + p)~,~~~
h_{R} = {R_{11}\over 8G_4}~(m-{l^2\over m})e^{-2\delta_0}= 
{1\over 2}(\epsilon-p) - {1\over n_1 n_2 n_3} J^2~.
\nonumber 
\ee
The space-time angular momentum is normalized so that $J$ is measured
in units of $\hbar$.  Thus, from semi-classical reasoning, we expect 
that $J$ is quantized as an integer. By introducing a single unit 
of angular momentum we see that the $h_R$ is quantized in units 
of $1/n_1 n_2 n_3$.

The angular momentum of the black hole breaks rotational invariance
of the background, so it is not guaranteed by symmetries that the 
near-horizon geometry contains a two-sphere $S^2$. In the present model
the linking of $AdS_3$ and $S^2$ is accomplished by the {\it global} 
features contained in the boundary conditions at infinity and
encoded in the coordinate shift (\ref{eq:shift}). 
It is therefore surprisingly simple to include angular momentum,
and thus more realism, while preserving full analytical control.
This makes the present model an attractive setting to study angular 
momentum. The precise value of the shift can be understood as 
follows: the potentials conjugate to the left- and right-moving 
string energies are:
\be
\beta_L = {\pi\over 2}\lambda^{3\over 2}m^{-{1\over 2}}e^{-\delta_0}~,~~~
\beta_R = {\pi\over 2}\lambda^{3\over 2}m^{1\over 2}
(m^2-l^2)^{-{1\over 2}}e^{\delta_0}~,
\nonumber
\ee
respectively; and the rotational velocity $\Omega$ is given through 
$\beta_H\Omega = {2\pi l\over\sqrt{m^2-l^2}}$,
where $\beta_H={1\over 2}(\beta_L+\beta_R)$ is the inverse of the 
Hawking temperature. 
Thus, in the ``co-moving'' frame where the
${\tilde\phi}$, given in (\ref{eq:shift}), is fixed,  
we have:
\be
\large({d\phi\over dt}\large)_{t=y,{\tilde\phi}} = 
{4l\over\sqrt{\lambda^3 m}}e^{-\delta_0}
= {\beta_H\Omega\over\beta_R}~,
\nonumber 
\ee
so the azimuthal angle $\phi$ is essentially shifted by the 
angular velocity $\Omega$. The factors of inverse temperatures and 
their significance for the wave functions of black hole perturbations 
are similar to the ones discussed for five-dimensional black holes 
in~\cite{cl98a}.

The direct connection between the near-horizon geometry and the 
underlying CFT appears to be valid for black holes in the near-extreme 
limit only.  Eventually, it will be important to test its validity and 
limitations away from the near-extreme limit. The structure indicated 
by angular momentum may play an important role in this 
endeavour~\cite{cy96b,fl97,bek97}. 

\vspace{0.2in} {\bf Acknowledgments:} 
MC would like to thank J. Polchinski and other participants 
in the duality program at ITP, Santa Barbara, for discussions.
This work is supported in part by DOE grant DOE-FG02-95ER40893.
Work at the ITP was further supported by the NSF under grant 
PHY94-07194.


\end{document}